# Metal-Dielectric-Graphene Sandwich for Surface Enhanced Raman Spectroscopy

Xuechao Yu[a], Jin Tao[a], Youde Shen[b], Guozhen Liang[a], Tao Liu[a], Yongzhe Zhang[c] and Qi Jie Wang[a,b,*]

[a] *School of Electrical and Electronic Engineering, Nanyang Technological University, 50 Nanyang Ave., 639798, Singapore*
[b] *Centre for Disruptive Photonic Technologies, Nanyang Technological University, 21 Nanyang Link, 63737, Singapore*
[c] *School of Renewable Energy and State Key Laboratory of Alternate Electrical Power System with Renewable Energy Sources, North China Electric Power University, Beijing 102206, China*

*Corresponding e-mail: qjwang@ntu.edu.sg*

**Raman intensity of Rhodamine B (RhB) is enhanced by inserting a thin high κ dielectric layer which reduces the surface plasmon damping at the gold-graphene interface. The results indicate that the Raman intensity increases sharply by plasmonic resonance enhancement while maintaining efficient fluorescence quenching with optimized dielectric layer thickness.**

## Introduction

Surface enhanced Raman spectroscopy (SERS) has attracted substantial interests because of its ability in providing structure information of chemical compounds or molecules at extremely low concentrations even at a single molecule level[1, 2]. It is widely accepted that there are three basic mechanisms, namely electromagnetic resonance (EM)[3], charge transfer resonance and molecule resonance[4,5] behind this Raman enhanced scattering phenomenon. One of the fascinating pursuits has been to detect the Raman signals of molecules adsorbed on a rough metal substrate that can provide hot spot sites where strong field enhancement occurs[6]. Metals and metal oxides with various morphologies and structures have been widely used as substrates for SERS[7-9]. However, the inevitable accompanying photoluminescence of metals and fluorescence emission of molecules is a major obstruction for high resolution detection and analysis, as it can boost the background of the Raman signal[10]. Thus one primary requirement of this technique is to quench the fluorescence background and increase the signal-to-noise ratio in the Raman spectrum.

Graphene, an atomic monolayer of hexagonally arranged carbon atoms, has been recently employed as a high efficiency quencher for fluorescence in SERS measurements because of the linear dispersion property of the electronic band near the Dirac point[11-14]. Although graphene itself is an attracting platform for SERS applications because of its enhancement of molecular adsorption[15], surface passivation[16] and its tunable Fermi level[17], the overall improvement on the signal-to-noise ratio in Raman spectrum is quite limited owing to the chemical mechanism which is determined by the electric field parallel to the surface of the metal substrate at the molecular binding site[18]. As a result, metal-on-graphene structure is explored based on the localized field enhancement properties of metals while graphene serves as a fluorescence quencher[19, 20]. However, when graphene is used as a substrate to support metals, the damping of plasmon on the graphene layer reduces the localized light intensity.

In this work, we deposit a thin $HfO_2$ layer between the graphene layer and gold nano-particles to reduce the plasmon damping process on graphene. The effect of this $HfO_2$ layer on the quenching efficiency of dye fluorescence is discussed. The results show that an optimized dielectric layer will obviously increase the localized light intensity and reduce the fluorescence emission. It is well known that Raman intensity of the molecules located near or between the metal nanoparticles can be enhanced by several orders of magnitude due to the remarkably enhanced localized electromagnetic field in the vicinity of metal nanostructures caused by localized surface plasmons resonance effect[6]. As demonstrated experimentally and theoretically, the electromagnetic field distribution is governed by the boundary conditions of the resonance; the particle size and distance are of great importance to the SERS activity [21]. In our design, graphene is employed as a fluorescence quencher, and gold nanoislands are simultaneously used to enhance the localized light intensity. Graphene is prepared by mechanical exfoliation of Kish graphite on $SiO_2$/Si wafer and characterized by Raman spectrum.

**Results and discussions**

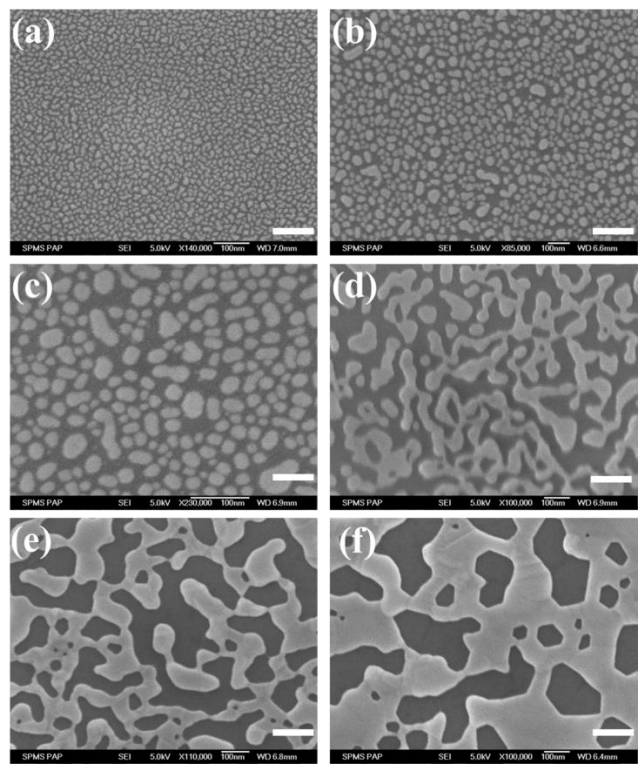

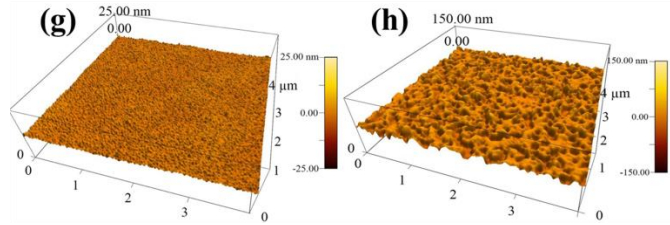

Fig. 1 SEM images of gold nanoislands with different sizes and shapes formed by an annealing process after the e-beam evaporation of gold films with different thicknesses (2 nm (a), 4 nm (b), 6 nm (c), 8 nm (d), 10 nm (e), 12 nm (f)) and AFM images of 8 nm gold on graphene before (g) and after (h) the annealing process and Raman spectrum of single layer graphene. The scale bars in (a-f) correspond to 100 nm.

In the next step, electron beam (e-beam) evaporation is employed to evaporate thin gold films, followed by an annealing process[22]. Fig. 1 (a)-(f) show that the evaporated Au layer is transformed into nanoislands on the surface of graphene with different sizes and shapes after the annealing process. The sizes and shapes of the nanoislands can be controlled by changing of the metal film thickness, the annealing temperature and time. In addition, atomic force microscopy (AFM) images in fig. 1 (g)-(h) also indicate that the gold films transfer to gold nanoislands with much larger surface roughness on graphene, which enhances the hot spot effect and thus a further enhancement in the local light intensity[6].

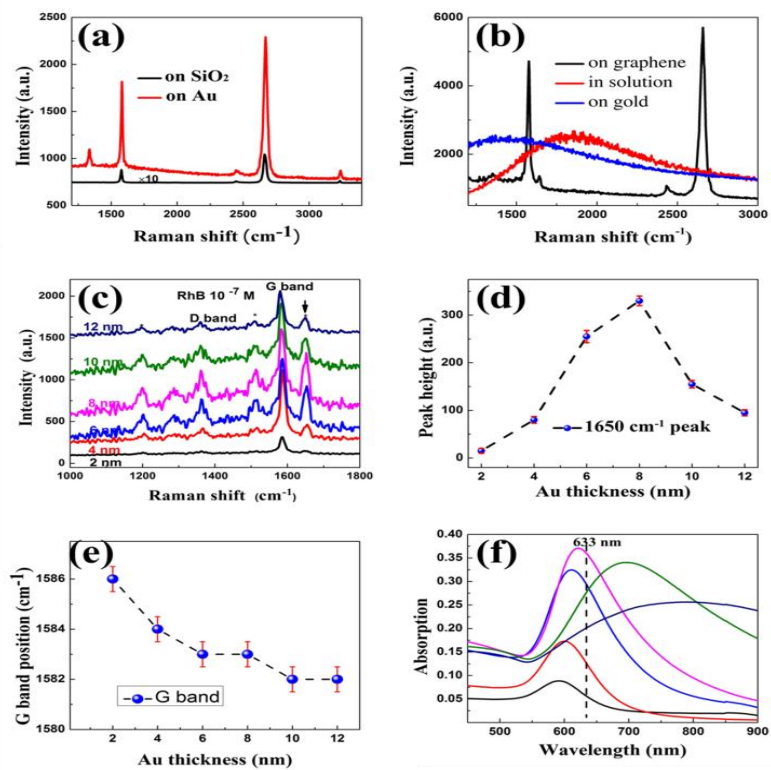

Fig. 2 Raman spectroscopy of single layer graphene with and without gold nanoislands (a), Raman spectroscopy of RhB on graphene, in solution and on gold film (b), Raman spectroscopy of RhB on annealed the gold-graphene substrate with different gold thicknesses (c); intensity of 1650 cm$^{-1}$ with different gold thicknesses (d), G-peak position of graphene with different thickness of gold after annealing process (e), absorption spectra of the gold-graphene substrate with different gold layer thickness after the annealing treatment(f), the colour lines in (f) correspond to the same thickness as in (c).

Although the light intensity was sharply enhanced by the gold nanoislands, there are no Raman signals for RhB when it absorbs on gold nanoislands as shown in fig. 2(b) since the Raman signals were submerged by the fluorescence as previous works reported[12]. Therefore, we employed graphene as a substrate to suppress fluorescence background for extracting Raman signals of low concentration RhB. The quality and Fermi level of graphene are of principal importance for fluorescence quenching process, which can be characterized from the Raman spectroscopies. From the Raman spectroscopy in Fig.2 (a) and (c), the G peak of graphene located at 1582 cm$^{-1}$, the Fermi level can be calculated from the following equation[23]:

$$\hbar \Delta \omega_G = \frac{\hbar A \langle D^2 \rangle_F}{\pi M \omega_0 (\hbar \upsilon_F)^2} |\Delta E_F|$$

, where $\langle D^2 \rangle_F$ is the deformation potential of the $E_{2g}$ mode[24], M is the atomic weight of carbon, $\omega_0$ is the frequency of the G-band in perfect graphene and $\upsilon_F$ is the Fermi velocity of graphene. The shift of Fermi level is around 0.2 eV, which is similar to previous reports where graphene-induced fluorescence quenching occurs[15, 17, 20, 25, 26]. On the other hand, the observable D peak, as shown in fig.2 (a) and (c), indicated that gold film deposition and annealing process induced defects in the graphene sheets.

We investigate the SERS spectrum of the gold-on-graphene substrate by measuring the Raman spectrum of Rhodamine B (RhB) adsorbed on the surface of graphene. Here, we define the intensity of the Raman signal to characterize the SERS effect, where $I_{signal} = I_{peak\ intensity} - I_{background}$. As shown in fig. 2 (c), the characteristic peaks of RhB are clearly observed at 1531 cm$^{-1}$, 1567 cm$^{-1}$ and 1650 cm$^{-1}$ with a concentration of 10$^{-7}$ M. It needs to be mentioned that after the gold deposition and the annealing process, the G peak of graphene shifted to 1586 cm$^{-1}$~1588 cm$^{-1}$ and the peak is split as indicated in fig. 2(e), which show that gold nanoislands deposition introduces n-type doping of graphene and the Fermi level is pushed down further[27], thus the fluorescence quenching efficiency can be enhanced. Throughout this work, we focus on the 1650 cm$^{-1}$ peak to characterize the Raman signal of RhB. Fig. 2 (d) shows that the samples with 8 nm gold film after annealing exhibit the highest Raman signal of RhB, which can be contributed to the size difference of the shrinked gold nano-islands after annealing. This is consistent with the measured absorption spectra of the same gold films on glass after the same treatment as shown in fig. 2(d) that the maximum enhancement occurs for a thickness of 8 nm. The enhancement can be verified by the absorption spectra of the same thickness of gold films deposited on glass after the same treatment as we fabricate the SERS device. As shown in fig. 2(f), the maximum absorption occurs for the samples with gold thickness of 8 nm after annealing.

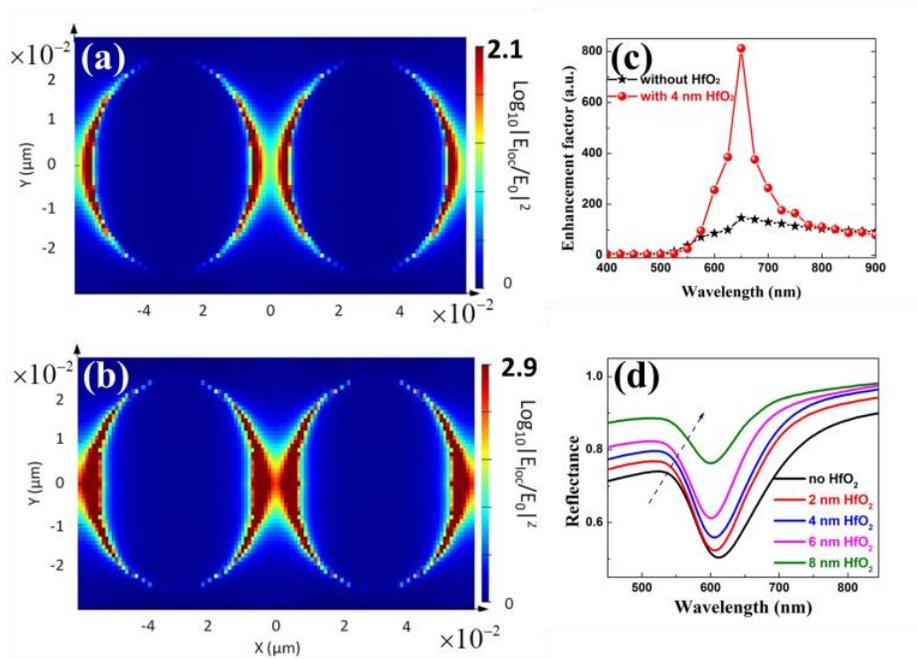

Fig. 3 Calculated intensity enhancement (top-view) of the electric filed with respect to the incident filed on gold nanoislands without (a) and with (b) $HfO_2$ layer and enhancement factor in the 400 nm-900 nm spectra with or without $HfO_2$ layer (c) calculated enhancement factor of gold-graphene substrate with and without a $HfO_2$ layer with a thickness of 4 nm; (d) measured reflection spectra of 8 nm gold on glass with different thicknesses of $HfO_2$ with the same annealing treatment of gold on graphene.

In order to further enhance the localized surface plasma resonance, a high-k, large band-gap dielectric layer is widely used to reduce the plasmon damping process and enhance plasmon resonance on the surface of metals[28, 29]. Fig. 3(a) and (b) show the simulation results of the normalized intensity enhancement (top-view) near the gold nanoislands without and with $HfO_2$ layer, respectively, sandwiched between the metal and graphene layers. Strong field enhancement appears on the surface of gold nanoislands and the gaps between two closely arranged nanoislands. The incident light wavelength used in the SERS experiment is at 633 nm. From the FDTD simulation results in fig. 3(c), one can see that the optical enhancement factor increases by about 4 times with an $HfO_2$ layer of 4 nm thick in a wavelength range from 600 nm to 700 nm. The intensity will be further improved for thicker $HfO_2$ layers. Furthermore, the resonance peak varies slightly for different thickness of $HfO_2$ as shown in fig. 3(d). As a result, we employ the optimized gold thickness of 8 nm for the following experiments to study the effects of varying the $HfO_2$ layer thickness on the Raman signal intensity.

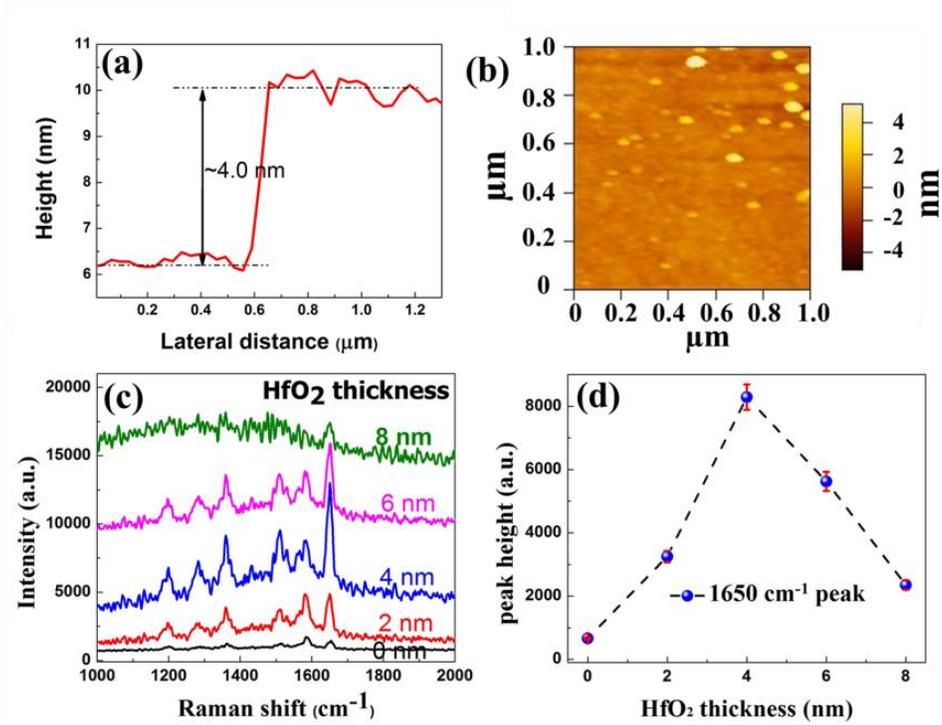

Fig. 4 AFM height profile (a) and image (b) of 40 ALD cycles HfO$_2$, Raman spectra of RhB on gold-HfO$_2$-graphene substrate with different thickness of HfO$_2$ (c), 1650 cm$^{-1}$ peak intensity dependence of HfO$_2$ thickness (d).

HfO$_2$ was deposited by atom layer deposition (ALD), and the thickness of HfO$_2$ can be well controlled by the ALD process. As all the HfO$_2$ films we fabricated are less than 10 nm, we reasonably assume that the growth rate is constant for all the samples (as seen in fig. 4 (a), 3.9~4.0 nm thick of HfO$_2$ corresponds to 40 ALD cycles, thus we get that the growth rate is around 0.1 nm per ALD cycle). Moreover, the surface roughness of the HfO$_2$ layer deposited by ALD is shown in fig. 3(b) that in a large area of 1 μm×1 μm, the film is continuous and quite uniform, only with little parts of higher lumps. The morphologies of gold particles on graphene and HfO2 are similar after annealing process (see SI). Like RhB in solution or on gold, the Raman signals were submerged in the fluorescence background as shown in fig. 2(b). As a result, we only treat HfO$_2$ layer in this structure as a blocker. Fig. 4(c) shows the experimental Raman spectrum of RhB on the gold-HfO$_2$-graphene substrate with various thicknesses of HfO$_2$ from 2-8 nm. G peak of graphene decreases gradually with the increase of HfO$_2$ thickness, which is attributed to the reduction of light penetrating through the dielectric layer as the reflectance increases dynamically as the thickness of HfO$_2$ layer increases as shown in fig. 3(d). However, the Raman signal of RhB for 4 nm HfO$_2$ is 2 times stronger than that for 2 nm HfO$_2$, indicating that the localized light intensity caused by gold nanoislands increases with the thickness of HfO$_2$ increases.

However, the Raman signals of RhB decrease rapidly when the thickness of HfO$_2$ is more than 4 nm because of the trade-off between fluorescence and Raman process. It is well known that for the two processes excited by light irradiation onto RhB, fluorescence is a linear process and Raman scattering is a non-linear process, respectively. Thus fluorescent signal of RhB is always the dominant one and several orders higher than the Raman scattering. In our design, for the metal-HfO$_2$-graphene structure, graphene is employed to quench the fluorescence emission so that we can get the Raman signal of RhB from its fluorescence background[25]. Even though HfO$_2$ could enhance the surface plasmon resonance on the gold island, it also hinders the quenching process of fluorescence as demonstrated theoretically[30-32] and experimentally[33] that the quenching rate, or the energy transfer rate, is proportional to $d^{-4}$ (d is the thickness of the dielectric layer). As shown in fig. 4 (d), we choose Raman peak intensity of 1650 cm$^{-1}$ as the characteristic peak of RhB and the peak intensity reaches its maximum at 4 nm thick of HfO$_2$.

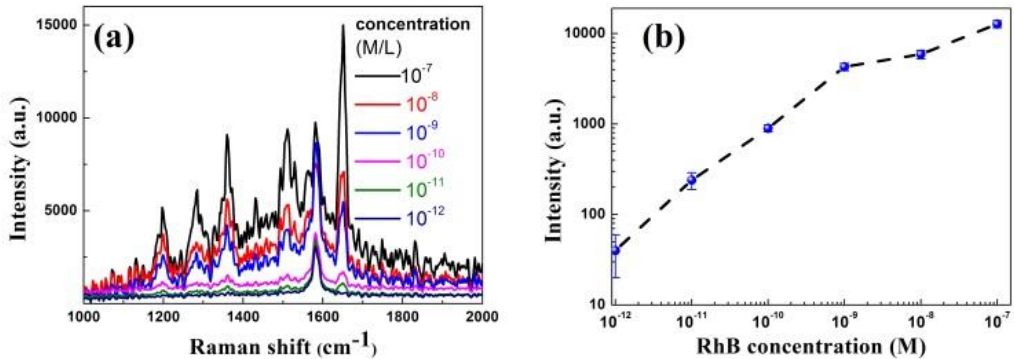

Fig. 5 (a) Raman spectra of RhB on the gold-HfO$_2$-graphene substrate with a 4 nm thick layer of HfO$_2$ and a 8 nm thick layer of Au after annealing with different concentration of RhB in methanol and (b) Raman signal dependence on the RhB concentration with the same structure (the inset is the molecular structure of RhB).

Furthermore, we employ the optimized structure with 8 nm thick of Au and 4 nm thick of HfO$_2$, in the SERS measurements for different RhB concentrations as shown in figure 5. It is clearly shown that the characteristic peak for RhB is obvious and detectable for the RhB solution with a concentration down to $10^{-12}$ M/L. Thus, such metal-dielectric-graphene structure enables picomolar detection sensitivity.

**Conclusions**

In summary, we demonstrate a novel gold-$HfO_2$-graphene structure that greatly improves the signal-to-noise ratio for the SERS effects. A high-k dielectric layer inserted between the metal and the graphene structure can largely enhance the surface plasmon effect and suppress the plasmon damping on the graphene layer as well. In addition, such structure also helps to quench the fluorescence emission of RhB, thus reduces the background noise in the Raman spectroscopy. This work provides a promising strategy and platform towards the design of SERS devices for achieving a high detectivity portable tool for molecular sensing for various applications, such as food safety inspection.

This work is supported by (MOE2011-T2-2-147 and MOE2011-T3-1-005) from Ministry of Education, Singapore.